\documentclass[manuscript]{aastex}

\usepackage{hyperref}
\usepackage{url}
\usepackage{lscape}
\usepackage{longtable}
\usepackage{latexsym}
\usepackage{graphics}
\usepackage{graphicx}
\usepackage{enumitem}
\usepackage{amsmath}
\usepackage{mathptmx}
\usepackage{latexsym}
\usepackage{fix-cm}
\usepackage{bigstrut}
\usepackage{booktabs}

\begin{document}

\title{Compact Relativistic Stars under Karmarkar Condition}

\author{B. S. Ratanpal\altaffilmark{1}}
\affil{Department of Applied Mathematics, Faculty of Technology \& Engineering, The Maharaja Sayajirao University of Baroda, Vadodara - 390 001, India}
\email{bharatratanpal@gmail.com}

\author{V. O. Thomas\altaffilmark{2}}
\affil{Department of Mathematics, Faculty of Science, The Maharaja Sayajirao University of Baroda,\\ Vadodara - 390 002, India}
\email{votmsu@gmail.com}

\and

\author{Rinkal Patel\altaffilmark{3}}
\affil{Department of Applied Science \& Humanities, Parul University, Limda, Vadodara - 391 760, India}
\email{rinkalpatel22@gmail.com}

\begin{abstract}

\noindent A class of new solutions for Einstein's field equations, by choosing the ansatz $e^{\lambda(r)}=\frac{1+ar^{2}}{1+br^{2}}$ for metric potential, are obtained under Karmarkar condition. It is found that a number of pulsars like 4U 1820-30, PSR J1903+327, 4U 1608-52, Vela X-1, PSR J1614-2230, Cen X-3 can be accomodated in this model. We have displayed the nature of physical parameters and energy conditions throughout the distribution using numerical and graphical methods for a particular pulsar 4U 1820-30 and found that the solution satisfies all physical requirements.

\end{abstract}


	\section{Introduction}
	\label{sec:1}
\noindent Eversince  \cite{Schwarzschild16} obtained the first solution of Einstein's field equations, a plethora of exact solutions are available at present, in literature. The interest in the study of anisotropic distrubutions has started with theoretical investigations of \cite{Ruderman72} and \cite{Canuto74} regarding the anisotropic nature of matter distribution in ultra-high densities. The impact of anisotropy on equilibrium of stellar configuration can be seen in the pioneering work of \cite{Bowers74}. \cite{Herrera97} have studied matter distribution incorporating anisotropy in pressure. A class of anisotropic solutions of spherically symmetric distribution of matter have been studied by \cite{Mak03}. \cite{maharaj06} have shown a procedure to generate anisotropic solutions from the known isotropic solutions. The impact of shear and electromagnetic field on stellar configuration has been studied by \cite{Sharma07}. 
	
\noindent A number of researchers have worked on spacetimes whose physical space obtained by putting $t=0$ has a definite geometry. \cite{vaidya82} have studied spherical distributions of matter on spacetime whose physical space has 3-spheroidal geometry. Charged analogue of this metric has been studied by \cite{Patel87}. \cite{Tikekar88} have obtained models of non-adiabatic gravitationally collapsing models with radial heat flux on the background of spheroidal spactime. The impact of anisotropy on \cite{vaidya82} model has been studied by \cite{karmarkar07}. 

\noindent \cite{Tikekar98} have studied relativistic models of stars on the background of pseudo-spheroidal spacetime and have shown that it can be used to describe equilibrium models of superdense stars. It has further shown that these models are stable under radial modes of pulsation. Non-adiabatic gravitational collapse of spherical stars incorporating radial heat flux have been studied by \cite{Thomas05} on the background of pseudo-spheroidal spacetime. \cite{Chattopadhyay10} have obtained the higher dimensional analogue of pseudo-spheroidal stellar models of \cite{Tikekar99}. \cite{Thomas15} have studied spherical distribution of matter by choosing a specific form for radial pressure on pseudo-spheroidal spacetime. \cite{Thomas16} have studied anisotropic models of superdense stars on the background of pseudo-spheroidal spacetime.
	
\noindent Another useful and geometrically significant spacetime widely used by researchers is the paraboloidal spacetime studied by  \cite{Jotania07}. \cite{tikekar09} have used this spacetime to obtain core-envelope models of superdense stars. Anisotropic models of stars on paraboloidal spacetime admitting quadratic equation of state have been studied by \cite{Sharma13}. New anisotropic solutions of relativistic star on paraboloidal spacetime has been obtained by \cite{ratanpal17}.  \cite{Thomas17} have obtained anisotropic compact star models with linear equation of state on the background of paraboloidal spacetime.
	
\noindent The embedding problems are geometrically significant problems in general theory of relativity. It was first studied by \cite{schlai71}. \cite{Nash56} proposed first isometric embedding theorem. The condition for embedding 4-dimensional spacetime in 5-dimensional Euclidean space was derived by \cite{Karmarkar48}. Such spacetimes are usually referred to a spacetimes of Class-I. The Karmarkar condition is given by
	\begin{equation}\label{karmarkar}
	R_{1414}R_{2323} = {R_{1212}R_{3434} + R_{1224}R_{1334}},   
	\end{equation}
	
\noindent \cite{pandey81} have found that for spherically symmetric spacetime metric to be of Class-I, it is further required that $ R_{2323}\neq 0 $ in
	(\ref{karmarkar}). Relativistic models of stars satisfying Karmarkar's condition have been extensively studied by \cite{maurya15}, \cite{gupta16}, \cite{maurya16}, \cite{smitha16}, \cite{bhar16}, \cite{maurya17}, \cite{maurya17b}.
	
\noindent In this article we have studied solutions of Einstein's field equations satisfying Karmarkar condition (\ref{karmarkar}) by choosing the metric pontential the ansatz $ e^{\lambda(r)}=\frac{1+ar^{2}}{1+br^{2}} $. If $ a=-\frac{k}{R^2} $ and $ b=-\frac{1}{R^2} $, the metric in Schwarzschild coordinates represents spheroidal spacetime metric proposed by \cite{vaidya82}. If $ a=\frac{k}{R^2} $
	and $ b=\frac{1}{R^2}$, the spacetime metric reduces to speudo-spheroidal spacetime metric considered by \cite{Tikekar98}. If we take $b = 0$ and $a=\frac{1}{R^2}$, the spacetime metric reduces to paraboloidal spacetime metric discussed by \cite{Jotania07}. 

\noindent We have organized the article as follows:
In section 2, we have given the Einstein's field equations and Karmarkar condition. The solution of Einstein field equations under Karmarkar condition is obtained in section 3. In section 4, physical plausibility conditions are described. The nature of various physical quantities throughout the distribution has been examined by taking a particular pulsar 4U 1820-30. In section 5, It has been concluded that a large variety of pulsars can be accomodated in this model incorporating Karmarkar condition.

\section{Einestein's field equations and Karmarkar condition}
\label{sec:2}
\noindent We consider the interior spacetime metric for static spherically symmetric fluid distribution as
\begin{equation}\label{IMetric}
ds^{2}=e^{\nu(r)}dt^{2}-e^{\lambda(r)}dr^{2}-r^{2}\left(d\theta^{2}+\sin^{2}\theta d\phi^{2} \right),
\end{equation}
with energy-momentum tensor
\begin{equation}\label{EMTensor}
T_{ij}=\left(\rho+p \right)u_{i}u_{j}-pg_{ij}+\pi_{ij},\;\;\;\;\;u^{i}u_{i}=1,
\end{equation}
where $\rho$ and $ p $  represent  density and isotropic fluid pressure respectively, $u^{i}$ is the  unit four velocity and anisotropic stress tensor $ \pi_{ij} $ is given by (\cite{MM89})
\begin{equation}
\pi_{ij}=\sqrt{3}S[c_{i}c_{j}-\frac{1}{3}(u_{i}u_{j}-g_{ij})],
\end{equation}
where $ S = S(r)  $ denotes the magnitude of anisotropy and $  c^{i}=(0,-e^{\lambda/2},0,0)$  denotes radially directed vector.
The non-vanishing components of  energy-momentum tensor are given by
\begin{equation}\label{5}
T_{0}^{0} = \rho ,\;\;\;\;\;\;   T_{1}^{1}= -\left(P+\frac{2S}{\sqrt{3}}\right), \;\;\;\;\;\;\;\;     T_{2}^{2} = T_{3}^{3} = -\left(P-\frac{S}{\sqrt{3}}\right).
\end{equation}
We shall denote
\begin{equation}
p_{r} = P+\frac{2S}{\sqrt{3}}     \;\;\;\;\;\;\;\;\;\;    p_{\perp} = P-\frac{S}{\sqrt{3}} ,
\end{equation}
and hence magnitude of anisotropy is given by
\begin{equation}\label{anisotropy}
S = \frac{p_{r}-p_{\perp}}{\sqrt{3}}.
\end{equation}
The Einstein's field equations, for spacetime metric (\ref{IMetric}) with energy-momentum tensor (\ref{EMTensor}), are given by
\begin{equation}\label{rho}
8\pi\rho=\frac{e^{-\lambda}\lambda'}{r}+\frac{1-e^{-\lambda}}{r^{2}} ,
\end{equation}
\begin{equation}\label{pr}
8\pi p_{r}=\frac{e^{-\lambda}\nu'}{r}+\frac{e^{-\lambda}-1}{r^{2}} ,
\end{equation}
\begin{equation}\label{pp}
8\pi p_{\perp}=e^{-\lambda} \left(\frac{\nu^{''}}{2} +\frac{\nu^2}{4}-\frac{\nu' \lambda'}{4}+\frac{\nu'-\lambda'}{2r}\right).
\end{equation}
The spacetime metric (\ref{IMetric}) is said to be of class-I type if it satisfies the Karmarkar condition (\ref{karmarkar}). The components of Riemann curvature tensor $R_{ijkl}$ for spacetime metric (\ref{IMetric}) are given by
\begin{equation*}
R_{2323}= r^{2}sin^{2}\theta \left(1-e^{-\lambda}\right),
\end{equation*}
\begin{equation*}
R_{1212}=\frac{1}{2}r\lambda^{'},
\end{equation*}
\begin{equation*}
R_{2424}=\frac{1}{2}r\nu^{'}e^{\nu}e^{-\lambda},
\end{equation*}
\begin{equation*}
R_{1224} =0,
\end{equation*}
\begin{equation*}
R_{1414}= e^{\nu}(\frac{\nu''}{2}+\frac{\nu'^2}{4}-\frac{\lambda'\nu'}{4}),
\end{equation*}
\begin{equation*}
R_{3434}=R_{2424}sin^{2}\theta.
\end{equation*}
The Karmarkar condition (\ref{karmarkar}) now takes the form
\begin{equation}\label{diifeqs}
\frac{\nu^{''}}{\nu^{'}}+\frac{\nu^{'}}{2}=\frac{\lambda^{'}e^{\lambda}}{2\left(e^{\lambda}-1\right)}.
\end{equation}
The general solution of equation (\ref{diifeqs}) is given by
\begin{equation}\label{enu}
e^{\nu}=\left[A+B\int\sqrt{(e^{\lambda(r)}-1)}dr\right]^{2},
\end{equation}
where A and B are constants of integration and $ e^{\lambda(r)} \neq 1 $. Using (\ref{pr}), (\ref{pp}),    (\ref{enu})	in (\ref{anisotropy}), the magnitude of anisotropy can be expressed in the form
\begin{equation}\label{S}
	8\pi\sqrt{3}S = -\frac{\nu'e^{-\lambda}}{4}\left[\frac{2}{r}-\frac{\lambda^{'}}{e^{\lambda}-1}\right]\left[\frac{\nu'e^{\nu}}{2rB^2}-1\right].
\end{equation}
In the case of isotropic distribution of matter, we have $S=0$ which leads to either $\frac{2}{r}-\frac{\lambda'}{e^{\lambda}-1}=0$ or $\frac{\nu' e^{\nu}}{2rB^2}-1=0$. The former case leads to \cite{Schwarzschild16} exterior solution and the latter gives the solution given by \cite{kohler65}.

\section{Anisotropic solution under Karmarkar condition}
	\label{sec:3}
\noindent The explicit expression for the potential $\nu$ can be obtained by choosing appropriate form for $\lambda$. We choose $e^{\lambda}$ in the form
\begin{equation}\label{e}
	e^{\lambda}=\frac{1+ar^{2}}{1+br^{2}},
\end{equation}
where a and b are constants. If $a=\frac{K}{R^{2}}$ and $b=1$, matric (\ref{IMetric}) represents the pseudo-spheroidal spacetime discussed by \cite{Tikekar98}. If $a=-\frac{K}{R^{2}}$ and $b=-1$ gives the \cite{vaidya82} spacetime and $a=1$, $b=0$ represent the paraboloidal spacetime studied by \cite{Jotania07}.

\noindent We shall assume here, that both $a$ and $b$ are not equal to zero. Substituting (\ref{e}) in (\ref{enu}), gives $e^{\nu}$ in the form
\begin{equation}\label{nu1}
	e^{\nu}=\left(A+B\frac{\sqrt{a-b}\sqrt{1+br^{2}}}{b}\right)^2.
\end{equation}
The spacetime metric (\ref{IMetric}) now takes the explicit form
\begin{equation}\label{IMetric1}
	ds^{2}=\left[A+B\frac{\sqrt{a-b}\sqrt{1+br^{2}}}{b}\right]^2 dt^{2}-\left(\frac{1+ar^{2}}{1+br^{2}}\right) dr^{2}-r^{2}\left(d\theta^{2}+\sin^{2}\theta d\phi^{2} \right).
\end{equation}
The expressions of matter density, radial pressure and tangential pressure are given by
\begin{equation}\label{rho1}
	8\pi\rho=\frac{(a-b)(3+ar^2)}{(1+ar^2)^{2}},
\end{equation}
\begin{equation}\label{pr1}
	8\pi p_{r}=\frac{Ab(b-a)+B\sqrt{a-b}\sqrt{1+br^2}(3b-a)}{(1+ar^2)(Ab+B\sqrt{a-b}\sqrt{1+br^2})},
\end{equation}
\begin{equation}\label{pp1}
	8\pi p_{\perp}=\frac{\sqrt{a-b}[-Ab\sqrt{a-b}+B\sqrt{1+br^2}(3b-a+abr^2)]}{(1+ar^2)^{2}(Ab+B\sqrt{a-b}\sqrt{1+br^2})}.
\end{equation}
The spacetime metric (\ref{IMetric1}) shoud match continuously with schwarzschild exterior metric
\begin{equation}\label{SEMetric}
	ds^{2}=\left(1-\frac{2M}{r}\right)dt^{2}- \left(1-\frac{2M}{r}\right)^{-1}dr^{2}-r^2(d\theta^2+sin^2\theta d\phi^2),
\end{equation}  
at the boundary of the star r = R. It leads to the following equations
\begin{equation}\label{M1}
	1-\frac{2M}{R}=\frac{1+bR^{2}}{1+aR^{2}},
\end{equation}
\begin{equation}\label{M2}
	\sqrt{1-\frac{2M}{R}}=A+B\frac{\sqrt{a-b}\sqrt{1+bR^2}}{b}.
\end{equation}
Further, the boundary condition $P_{r}\left(r=R\right)=0$ gives
\begin{equation}\label{M3}
	Ab\sqrt{a-b}=B\sqrt{1+bR^2}\left(3b-a\right).
\end{equation}
Equations (\ref{M1}), (\ref{M2}) and (\ref{M3}) determine the constants $A$, $B$ and the total mass enclosed inside the radius R as 
\begin{equation}\label{A}
	A=\frac{(3b-a)\sqrt{1+bR^2}}{2b\sqrt{1+aR^2}},
\end{equation}
\begin{equation}\label{B}
	B = \frac{\sqrt{a-b}}{2\sqrt{1+ar^2}},
\end{equation}
\begin{equation}\label{Mass}
	M = \frac{\left(a-b\right)R^{3}}{2\left(1+aR^2 \right)}.
\end{equation}
Equations (\ref{rho1}) through (\ref{pp1}) now take the form
\begin{equation}\label{rho2}
	8\pi\rho=\frac{(a-b)(3+ar^2)}{(1+ar^2)^{2}},
\end{equation}
\begin{equation}\label{pr2}
	p_{r}=\frac{\left(3b-a\right)\left(b-a \right)}{1+ar^{2}}\left[\frac{\sqrt{1+bR^{2}}-\sqrt{1+br^{2}}}{\left(3b-a \right)\sqrt{1+bR^{2}}-\left(b-a \right)\sqrt{1+br^2}} \right],
\end{equation}
\begin{equation}\label{pp2}
	p_{\perp}=\frac{-(a-b)[3b(1+br^2-\sqrt{1+br^2}\sqrt{1+bR^2})+a(-1+b^2r^4+\sqrt{1+br^2}\sqrt{1+bR^2})]}{(1+ar^2)^{2}\sqrt{1+br^2}[b(\sqrt{1+br^2}-3\sqrt{1+bR^2})+a(-\sqrt{1+br^2}+\sqrt{1+bR^2})]}.
\end{equation}
The expression for anisotropy (\ref{S}) can be explicitly written as
\begin{equation}\label{ani2}
	8\pi\sqrt{3}S = \frac{a(a-b)r^2[a(1+br^2-\sqrt{1+br^2}\sqrt{1+bR^2})+b(-2-2br^2+3\sqrt{1+br^2}\sqrt{1+bR^2})]}{(1+ar^2)^{2}\sqrt{1+br^2}[b(\sqrt{1+br^2}-3\sqrt{1+bR^2})+a(-\sqrt{1+br^2}+\sqrt{1+bR^2})]}.
\end{equation}
It can be noticed that anisotropy of the distribution is zero at the centre of the star.	
	
\section{Physical Plausibility Conditions}
	\label{sec:4}
\noindent A physically acceptable stellar model should comply with the following conditions throughout its region of validity.
	\\ $ (i) $\;\;\;\;   $ \rho(r) \ge 0 ,\;\;\;\;p_{r}(r)\ge 0 ,\;\;\;\;p_{\perp}(r)\ge 0 $   \;\;\;\;\;\;\;for   $ 0 \le r \le R  $
	\\ $(ii) $    \;\;\;\; $\frac{d\rho}{dr}\le 0 ,\;\;\;\;  \frac{dp_{r}}{dr} \le 0 ,\;\;\;\;   \frac{dp_{\perp}}{dr}\le 0  $ \;\;\;\;\;\; for  $ 0 \le r \le R  $
	\\ $ (iii) $\;\;\;\; $0 < \frac{dp_{r}}{d\rho} < 1$ ,\;\;\;\; $0 < \frac{dp_{\perp}}{d\rho} < 1 $ \;\;\;for $ 0 \le r \le R  $  
	\\  $(iv) $\;\;\;\;  $\rho-p_{r}-2p_{\perp}\ge 0 $ \;\;\;\; for $0 \le r \le R  $ 
	\\$(v) $\;\;\;\;  $\Gamma > \frac{4}{3}  $ ,\;\;\;\;\;\;\; for $ 0 \le r \le R  $ 
	
	\begin{table}[h]
		\caption{Estimated physical values of parameters based on the observational data}
		\label{tab:1}
		\begin{tabular}{llllll}
			\hline\noalign{\smallskip}
			\textbf{STAR} &  {$ \mathbf{M} $} & {$ \mathbf{R} $} & {$ \mathbf{\rho_c} $} & {$ \mathbf{\rho_R} $} & {$ \mathbf{u (=\frac{M}{R})} $} \\
			&  $ \mathbf{(M_\odot)} $ & $ \mathbf{(Km)} $ & \textbf{(MeV fm{$\mathbf{^{-3}}$})} & \textbf{(MeV fm{$\mathbf{^{-3}}$})}   \\
			\noalign{\smallskip}\hline\noalign{\smallskip}
			\textbf{4U 1820-30} 	  & 1.25   & 9.1   &  804.032  & 309.128   & 0.137 \\
			\textbf{PSR J1903+327} 	  & 1.35  & 9.438  &  804.032  & 293.779   & 0.142 \\
			\textbf{4U 1608-52} 	  & 1.31  & 9.31   &  804.032  & 299.495   & 0.140 \\
			\textbf{Vela X-1} 	      & 1.38  & 9.56   &  804.032  & 288.439   & 0.144 \\
			\textbf{PSR J1614-2230}   & 1.42  & 9.69   &  804.032  & 282.863   & 0.146 \\
			\textbf{Cen X-3}          & 1.27  & 9.178  &  804.032  & 305.513   & 0.138 \\
			\noalign{\smallskip}\hline
		\end{tabular} 
	\end{table}
\noindent We shall use the above conditions to find the bounds on the model parameters a and b. Density $ \rho $ is positive and decreasing throughout
	the distribution if $ a>b. $ The radial pressure $ p_{r} $ is positive and decreasing throughout
	the distribution if $ a\le \frac{4}{R^2} $ and $ a>3b. $ The transverse pressure $ p_{\perp} $ is positive and decreasing throughout
	the distribution if $ \frac{0.2749}{R^2} < a < \frac{4}{R^2}. $ The conditions $0< \frac{dpr}{d\rho} < 1 $ and  $0< \frac{dp\perp}{d\rho} < 1 $ impose the restrictions $a \le \frac{2.7847}{R^2}$  and $\frac{0.4384}{R^2} < a \le \frac{2.7025}{R^2} $, $\rho-p_{r}-2p_{\perp}\geq 0$ is satisfied if $ 0< a \le\frac{2}{R^2} $ and $ a>b. $ The adiabatic index $ \Gamma > \frac{4}{3}$ if $ a\le \frac{0.8202}{R^2}. $
Thus the conditions (i) through (v) are satisfied if 
\begin{equation}\label{value}
	\frac{0.4384}{R^2} < a \le \frac{0.8202}{R^2},\; a>3b.
\end{equation}
\noindent We shall examine the viability of the present model to represent some well-known pulsars whose mass and size are known.

\begin{figure}[ht]
	\includegraphics[scale = 1.25]{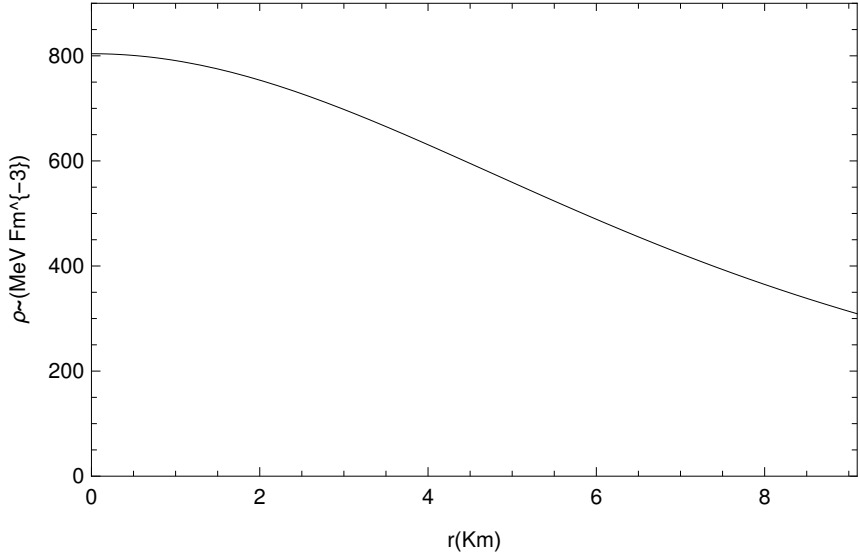} 
	\caption{Variation of density against radial variable $r$. 
		\label{fig:1}}
\end{figure}
\begin{figure}[ht]
	\includegraphics[scale = 1.25]{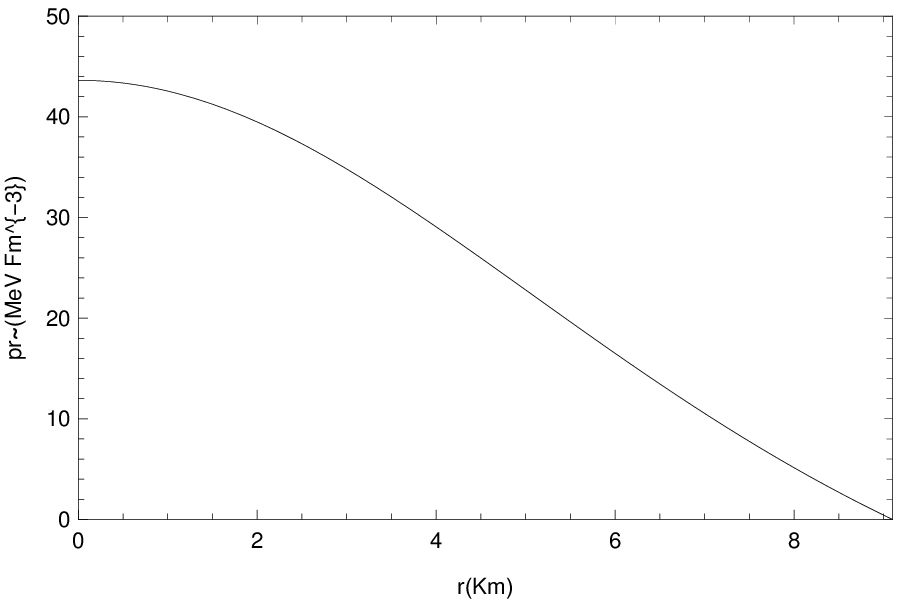}
	\caption{Variation of radial pressures against radial variable $r$.
		\label{fig:2}}
\end{figure}

\begin{figure}[ht]
	\includegraphics[scale = 1.25]{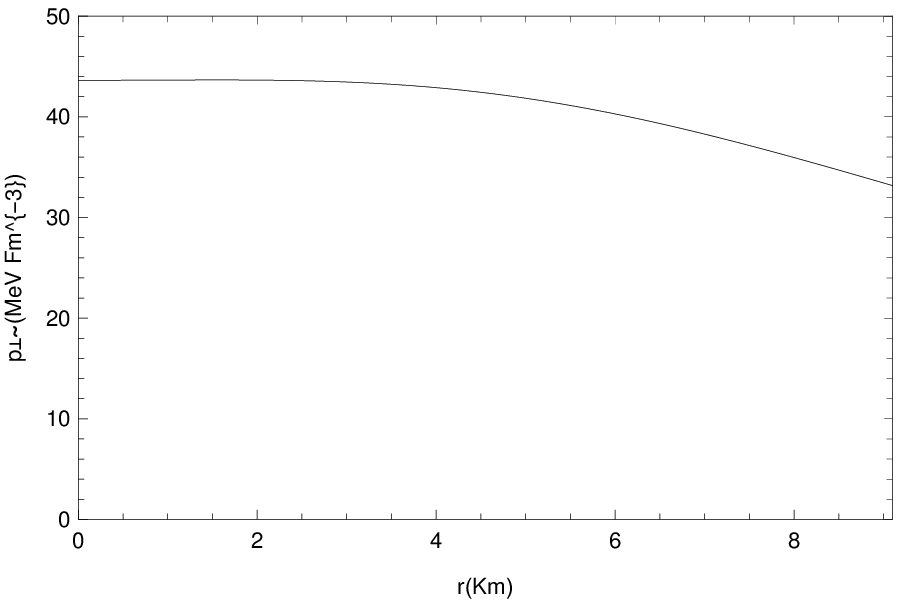}
	\caption{Variation of transverse pressures against radial variable $r$ 
		\label{fig:3}}
\end{figure}

\begin{figure}[ht]
	\includegraphics[scale = 1.25]{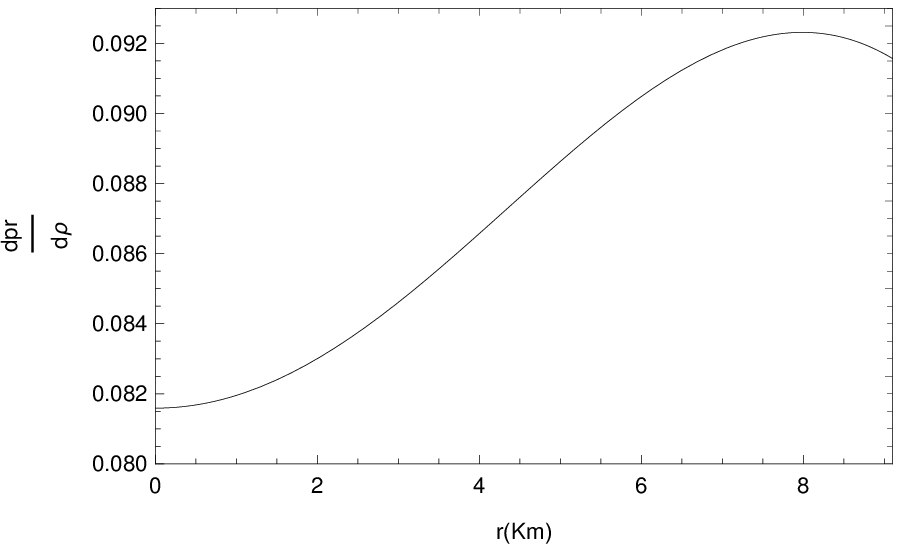}
	\caption{Variation of $ \frac{dp_r}{d\rho} $ against radial variable $r$. 
		\label{fig:4}}
\end{figure}

\begin{figure}[ht]
	\includegraphics[scale = 1.25]{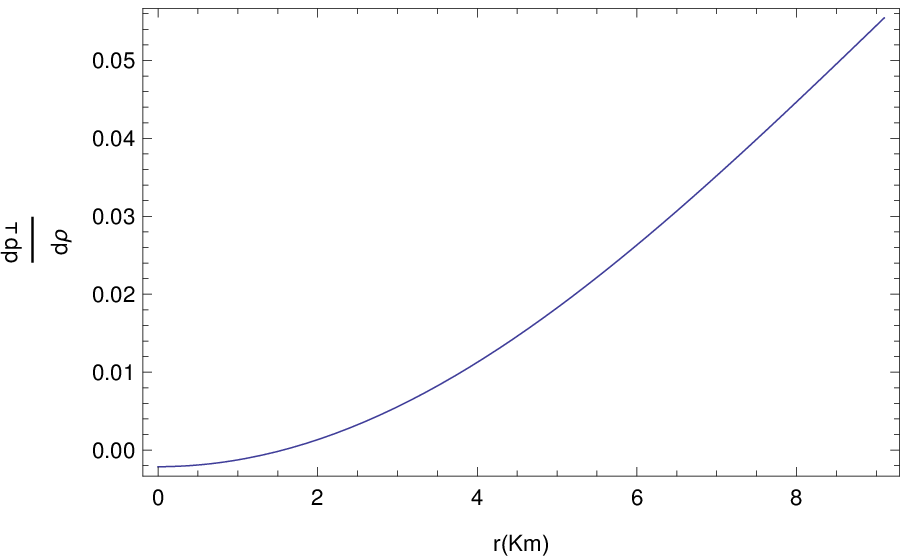}
	\caption{Variation of $ \frac{dp_\perp}{d\rho} $ against radial variable $r$.
		\label{fig:5}}
\end{figure}

\begin{figure}[ht]
	\includegraphics[scale = 1.25]{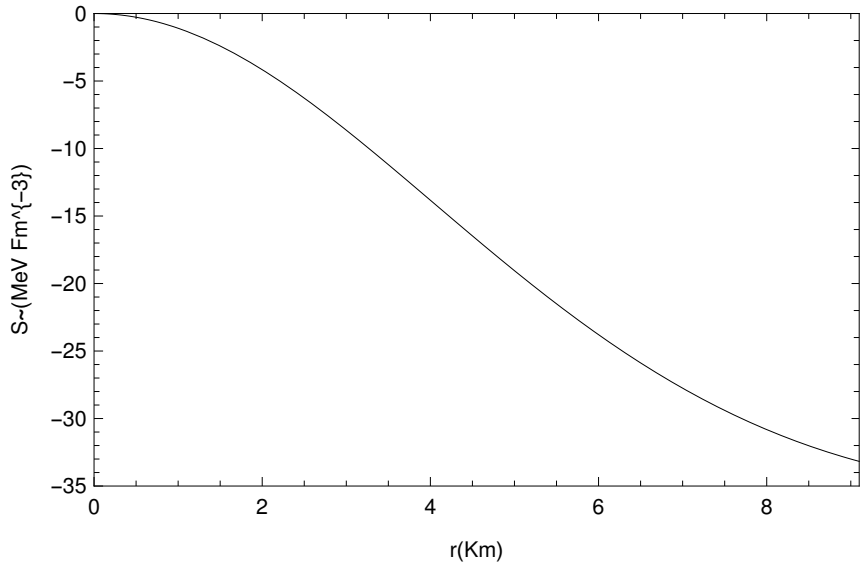}
	\caption{Variation of anisotropies against radial variable $r$. 
		\label{fig:6}}
\end{figure}

\begin{figure}[ht]
	\includegraphics[scale = 1.25]{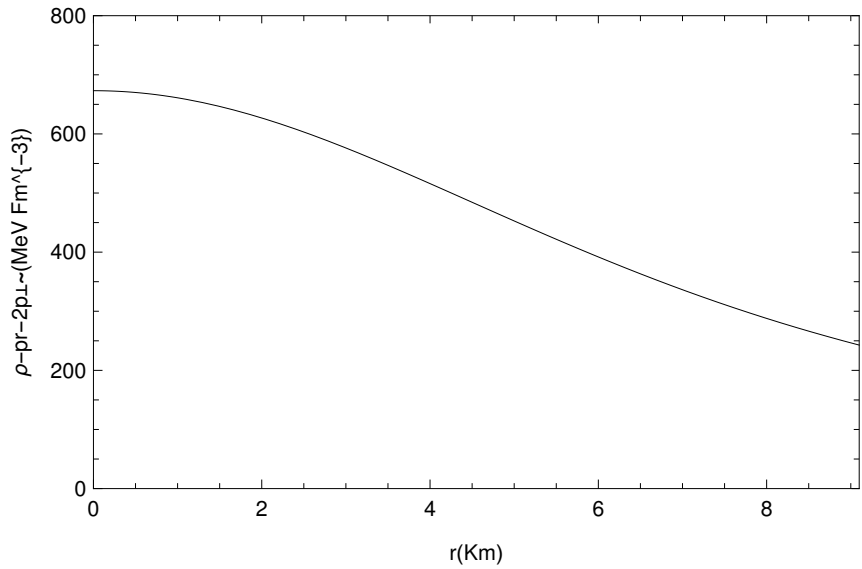}
	\caption{Variation of strong energy condition against radial variable $r$. 
		\label{fig:7}}
\end{figure}

\begin{figure}[ht]
	\includegraphics[scale = 1.25]{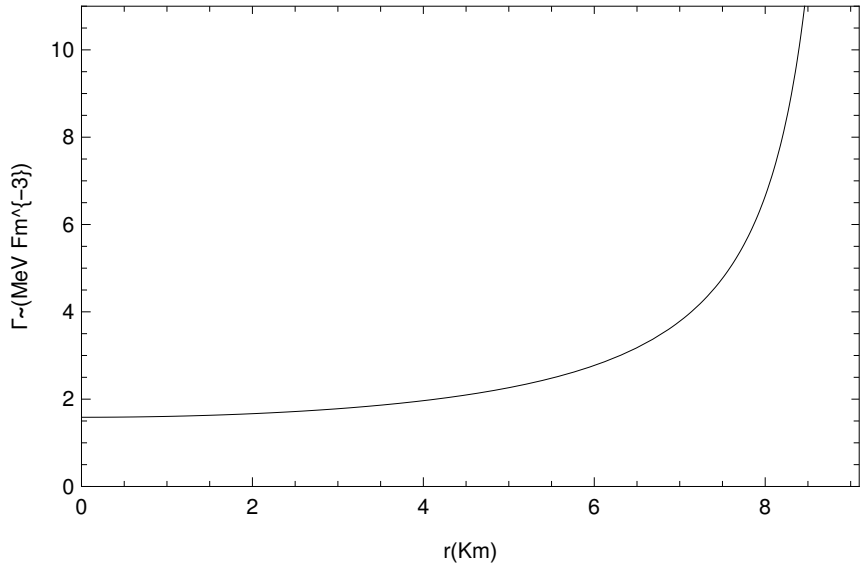}
	\caption{Variation of adiabatic Index against radial variable $r$. 
		\label{fig:8}}
\end{figure}

\begin{figure}[ht]
	\includegraphics[scale = 1.25]{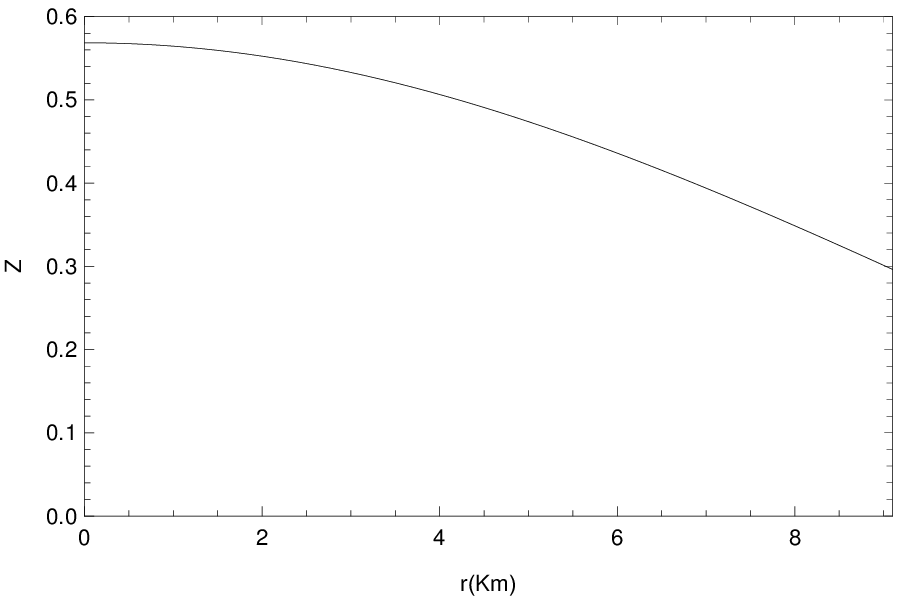}
	\caption{Variation of surface redshift against radial variable $r$. 
		\label{fig:9}}
\end{figure}
\section{Discussion}
\noindent We have used the present model to a large variety of compact stars like 4U 1820-30, PSR J1903+327, 4U 1608-52, Vela X-1, PSR J1614-2230, Cen X-3, whose masses and radii are known (\cite{Gango13}). The central and surface densities are calculated and displayed in Table 1 along with the compactification factor. We have shown that energy and stability conditions are satisfied for all pulsars listed in Table 1 for suitable bounds for the parameters $a$ and $b$ depending on the radii of different pulsars under consideration. Due to the complexity of expressions involved, it is difficult to examine the physical acceptability conditions analytically. Hence we have adopted graphical method.

\noindent In order to examine the nature of physical quantities throughout the distribution, we have considered the pulsar 4U 1820-30 whose estimated mass is $M=1.25\; M_{\odot}$ and radius $R=9.1\; km$. The expressions (\ref{value}) now take the form
\begin{equation}\label{Bound1}
	0.0053<a\leq 0.0099,\;a>3b.
\end{equation} 
We have taken the value of $a$ as the upper bound $0.0099$, $b=0.001$ and examined the physical, energy and stability conditions of the pulsar throughout its region of validity.

\noindent In Fig. 1 we have shown the variation of density for $0\leq r \leq 9.1$. It is clear from the graph that the density is a decreasing function of $r$. In Fig. 2 and Fig. 3 we have shown the variation of radial and transverse pressure throughout the star. It can be seen that both pressures are decreasing radially outward. In Fig. 4 and Fig. 5 we have displayed the variation of $\frac{dp_{r}}{d\rho}$ and $\frac{dp_{\perp}}{d\rho}$ against $r$. Both quantities satisfy the restriction $0<\frac{dp_{r}}{d\rho}<1$ and $0<\frac{dp_{\perp}}{d\rho}<1$ indicating that the sound speed throughout the star is less than the speed of light.

\noindent The variation of anisotropy is showin in Fig. 6. It can be noticed that anisotropy vanishes at the centre and decreases towards the boundary. Fig. 7 indicates that the strong energy condition $\rho-p_{r}-2p_{\perp}>0$ is satisfied througout the distribution. In order that a relativistic equilibrium model of a compact star is stable model, the adiabatic index $\Gamma=\frac{\rho+p_{r}}{p_{r}}\frac{dp_{r}}{d\rho}>\frac{4}{3}$ throughout the distribution. Fig. 8 indicates that the condition $\Gamma>\frac{4}{3}$ is satisfied in the region $0\leq r\leq 9.1$. In Fig. 9, we have shown that redshift $z$ is less than 1 and decreases radially outward.

\noindent It has been concluded that a large number of pulsars with known masses and radii can be accomodated in the present model satisying Karmarkar condition.
\pagebreak

\section*{Acknowledgement}
    BSR would like to thank IUCAA, Pune for the facilities and hospitality provided to him where the part of work was carried out.

\end{document}